\documentclass[twocolumn,aps,prl,longbibliography]{revtex4-1}
\usepackage{times}
\usepackage{graphicx}
\usepackage{amsfonts}
\usepackage{amsmath, amsthm, amssymb}
\usepackage{dsfont}
\usepackage{color}
\usepackage{empheq}
\usepackage{standalone}
\usepackage[most]{tcolorbox}
\usepackage{physics}
\usepackage{bm}
\usepackage[colorlinks=true, allcolors=black, citecolor=blue, linkcolor=blue]{hyperref}
\usepackage[utf8]{inputenc}
\usepackage[margin=1.0in]{geometry}
\usepackage{amssymb,amsmath,amsfonts,amscd}
\usepackage{mathrsfs}
\usepackage{siunitx}
\usepackage{commath}
\usepackage{graphicx}       
\graphicspath{ {Images/} }  
\usepackage{braket}
\usepackage{empheq}
\usepackage[most]{tcolorbox}
\usepackage{physics}

\usepackage{bibunits}
\defaultbibliography{Refs}
\defaultbibliographystyle{apsrev4-1}

\usepackage{hyperref} 
\hypersetup{
    colorlinks=true,
    linkcolor=black,    
    citecolor=black,    
    urlcolor=blue,      
}
\renewcommand{\rm}[1]{\mathrm{#1}}

\renewcommand{\v}[1]{{\bf{#1}}}

\usepackage{soul}

\allowdisplaybreaks[1]

\begin{document}

\title{Enhancement of superconductivity mediated by antiferromagnetic squeezed magnons}

\author{Eirik Erlandsen}
\affiliation{\mbox{Center for Quantum Spintronics, Department of Physics, Norwegian University of Science and Technology,}\\NO-7491 Trondheim, Norway}

\author{Akashdeep Kamra}
\affiliation{\mbox{Center for Quantum Spintronics, Department of Physics, Norwegian University of Science and Technology,}\\NO-7491 Trondheim, Norway}

\author{Arne Brataas}
\affiliation{\mbox{Center for Quantum Spintronics, Department of Physics, Norwegian University of Science and Technology,}\\NO-7491 Trondheim, Norway}

\author{Asle Sudb{\o}}
\email[Corresponding author: ]{asle.sudbo@ntnu.no}
\affiliation{\mbox{Center for Quantum Spintronics, Department of Physics, Norwegian University of Science and Technology,}\\NO-7491 Trondheim, Norway}

\begin{abstract}
We investigate theoretically magnon-mediated superconductivity in a heterostructure consisting of a normal metal and a two-sublattice antiferromagnetic insulator. The attractive electron-electron pairing interaction is caused by an interfacial exchange coupling with magnons residing in the antiferromagnet, resulting in p-wave, spin-triplet superconductivity in the normal metal. Our main finding is that one may significantly enhance the superconducting critical temperature by coupling the normal metal to only one of the two antiferromagnetic sublattices employing, for example, an uncompensated interface. Employing realistic material parameters, the critical temperature increases from vanishingly small values to values significantly larger than 1 K as the interfacial coupling becomes strongly sublattice-asymmetric. We provide a general physical picture of this enhancement mechanism based on the notion of squeezed bosonic eigenmodes.
\end{abstract}

\maketitle


\textit{Introduction}. -- Hybrids comprised of a magnetic insulator coupled to a conducting layer allow for interconversion between magnonic and electronic spin currents~\cite{Tserkovnyak2002,Kajiwara2010,Chumak2015,Weiler2013,Zhang2012,Adachi2013,Takahashi2010,Cornelissen2015,Goennenwein2015,Maekawa2012,Bauer2012}. Spin Hall effect~\cite{Hirsch1999,Sinova2015} in the conductor has further been exploited to electrically control and detect the magnonic spin currents~\cite{Saitoh2006}, thereby enabling their integration with conventional electronics. The ensuing newly gained control over spin currents has instigated a wide range of magnon transport based concepts and devices~\cite{Cornelissen2015,Goennenwein2015,Zhang2012,Chumak2014,Kruglyak2010,Ganzhorn2016,Lebrun2018}. Conversely, magnons in the magnet can mediate electron-electron attraction in the conducting layer. The resulting magnon-mediated superconductivity has been investigated  theoretically in normal metals~\cite{Rohling2018,Fjaerbu2019} as well as topological insulators~\cite{Kargarian2016,Hugdal2018}, and experimentally~\cite{Gong2017}. Magnon-mediated indirect exciton condensation has also been predicted recently~\cite{Johansen2019}.

Interest in antiferromagnets (AFMs) has recently been invigorated~\cite{Jungwirth2016,Baltz2016,Gomonay2018,Gomonay2018} due to their distinct advantages over ferromagnets (FMs), such as minimization of stray fields, sensitivity to external magnetic noise, and low-energy magnons. The demonstration of electrically-accessible memory cells based on AFMs~\cite{Wadley2016,Kosub2017} and spin transport across micrometers~\cite{Lebrun2018} corroborates their high application potential. Furthermore, their two-sublattice nature allows for unique phenomena~\cite{Kamra2018B,Ohnuma2013}, such as topological spintronics~\cite{Libor2018} and strong quantum fluctuations, not accommodated by FMs. AFMs with uncompensated interfaces, proven instrumental in exchange biasing~\cite{Nogues1999,Nogues2005,Stamps2000,Zhang2016,Manna2014,He2010,Belashchenko2010} FMs for contemporary memory technology, have been predicted to amplify spin transfer to an adjacent conductor~\cite{Kamra2017B}. Recently, a theoretical proposal for proximity-inducing spin splitting in a superconductor using an uncompensated AFM, along with an experimental feasibility study based on existing literature, has also been put forward~\cite{Kamra2018}. 

\begin{figure}[b]
\includegraphics[width=0.90\columnwidth,trim= 0.1cm 0.1cm 0.0cm 0.1cm,clip=true]{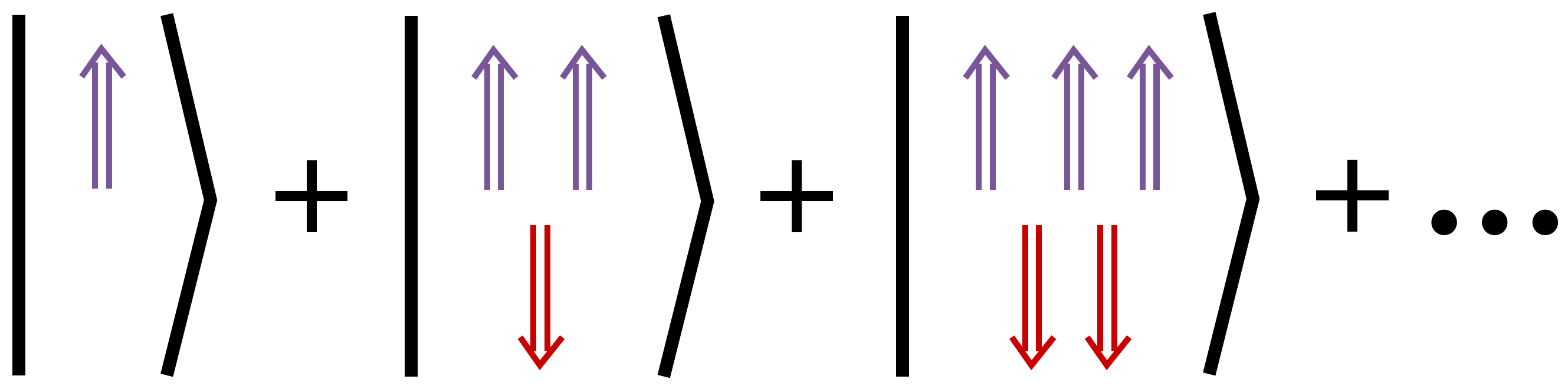}
\caption{Representation of a spin-up antiferromagnetic squeezed magnon~\cite{Kamra2019}. The squeezed excitation is a coherent superposition of states with N+1 spin-up and N spin-down magnons. Each of the constituent states possesses unit net spin, but varies in its spin content on each sublattice thereby resulting in strong quantum fluctuations.}
    \label{fig:sqmagnon}
\end{figure}

 \begin{figure*}[t]
 \includegraphics[width=1.87\columnwidth,trim= 0.0cm 0.0cm 0.0cm 0.0cm,clip=true]{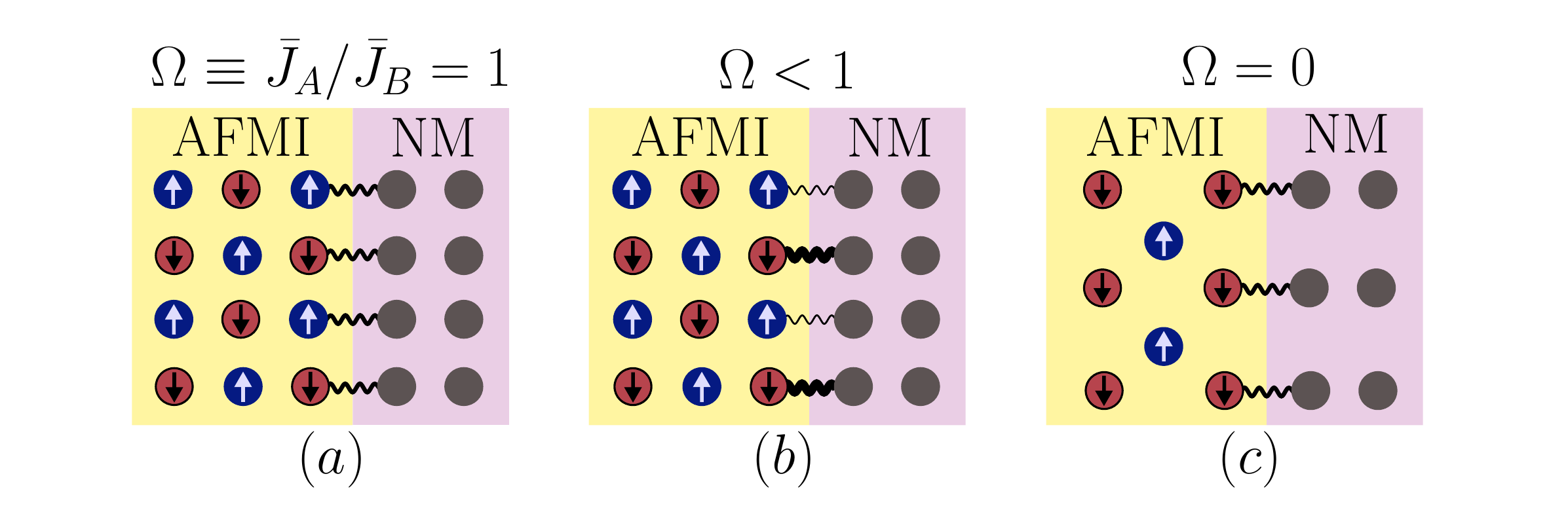}
\caption{Schematic depiction of the antiferromagnetic insulator (AFMI)/normal metal (NM) bilayer including its sublattice-asymmetric interfacial exchange coupling. Panels (a) and (c), respectively, illustrate compensated and uncompensated interfaces that are realized in experiments, for example via AFMI layer growth in a specific crystal orientation. Panel (b) shows the model employed in our analysis, which conveniently allows capturing the full range of interfacial asymmetry via the parameter $\Omega$.}
     \label{fig:System}
 \end{figure*}

Within the standard theory of boson-mediated superconductivity~\cite{PhysRev.108.1175,Schrieffer1999}, the superconducting critical temperature $T_c$ is determined by an energy scale set by some high-frequency cutoff $\omega_c$ on the boson-spectrum, coupling of electrons with these bosons, and the single-particle electronic density of states on the Fermi-surface. The latter two combine to an effective dimensionless coupling constant $\lambda$. In the simplest case, $T_c \sim \omega_c \exp(-1/\lambda)$. An enhancement of electron-phonon coupling, and thus $\lambda$, possibly due to a feedback loop involving strong correlation effects, typically results in a larger $T_c$~\cite{He2018}. An increase of $\lambda$ caused by nonequilibrium squeezing~\cite{Gerry2004} of phonons has been suggested~\cite{Knap2016,Babadi2017} as a mechanism underlying experimentally observed, light-induced transient enhancement of $T_c$ in some superconductors~\cite{Fausti2011,Mankowsky2014,Mitrano2016}. 

\begin{center}
	\begin{figure}[t]
		\includegraphics[width=0.88\columnwidth,trim= 8.5cm 8.8cm -1.5cm 0.8cm,clip=true]{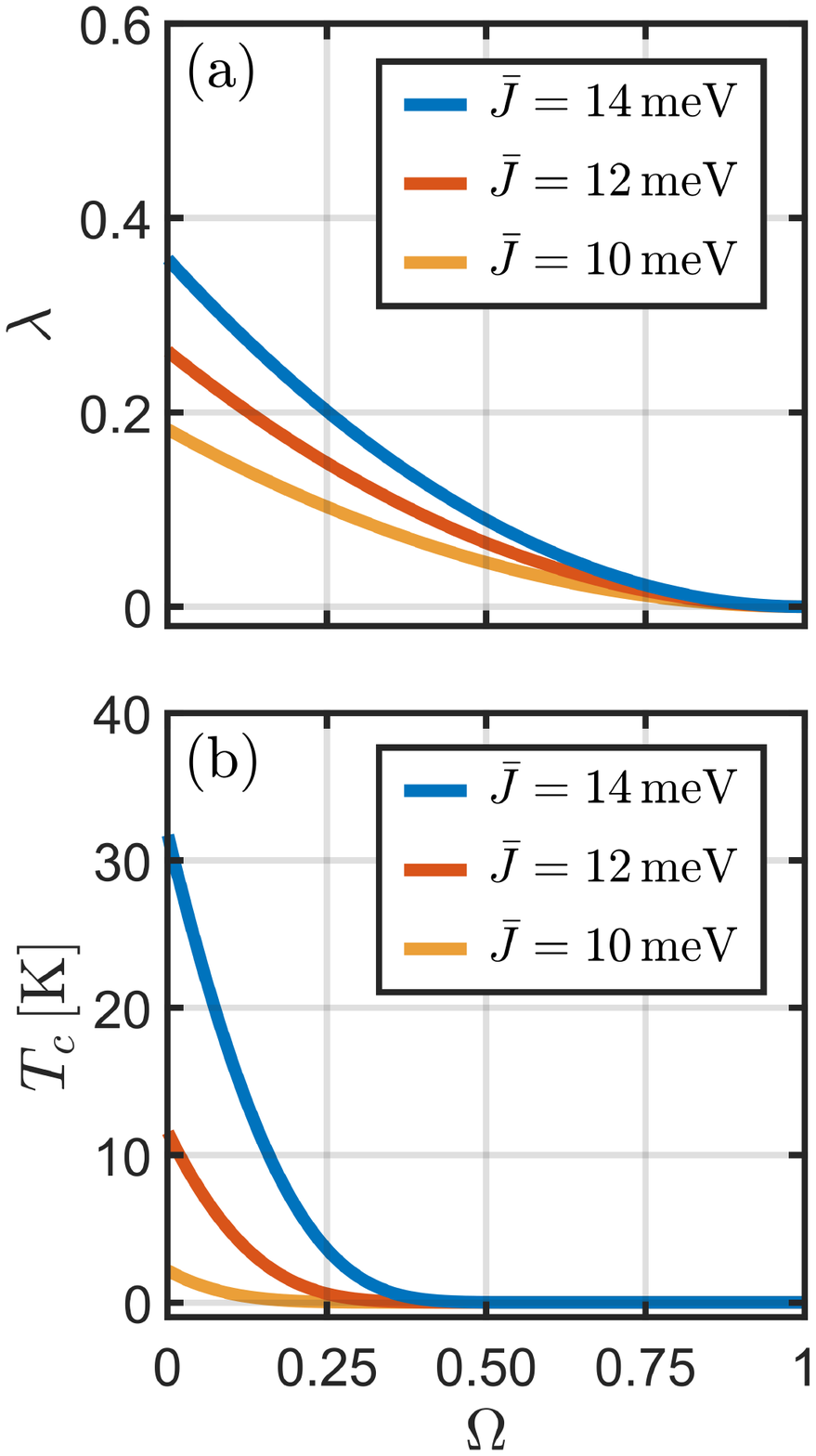}
		\caption{(a) Dimensionless coupling constant $\lambda$ and (b) superconducting critical temperature $T_c$ vs. coupling asymmetry parameter $\Omega \equiv \bar{J}_A/\bar{J}_B$, with $\bar{J} \equiv \bar{J}_B$. $\Omega = 1$ and $0$, respectively, correspond to compensated and uncompensated antiferromagnetic interfaces (Fig. \ref{fig:System}). An increase in both $\lambda$ and $T_c$ in the latter case constitutes our main result.}
		\label{fig:Tc}
	\end{figure}
\end{center}

In this Letter, we theoretically demonstrate a drastically increased, attractive, magnon-mediated electron-electron (e-e) interaction, exploiting the two-sublattice nature of, and equilibrium squeezing-mediated strong quantum fluctuations (Fig.~\ref{fig:sqmagnon}) in, an AFM~\cite{Kamra2017,Kamra2019}. We study a bilayer structure in which a normal metal (NM) exchange couples equally or differently to the two sublattices of an AFM insulator (AFMI), as depicted in Fig.\,\,\ref{fig:System}, and find a significant enhancement of the attractive e-e interaction in the latter case. This is attributed to an amplification of the electron-magnon coupling that appears through magnon coherence factors acting constructively, instead of destructively as they do in the case of equal coupling to both sublattices. The resulting increase in $\lambda$ produces a significant enhancement in $T_c$ with sublattice-symmetry-breaking of the interfacial exchange coupling between the NM and AFMI (Fig.~\ref{fig:Tc}). We also comment on the experimental feasibility and optimal materials for realization of the predicted p-wave, spin-triplet superconducting state in these engineered bilayers.
 
A physical picture of this electron-magnon coupling enhancement, detailed elsewhere~\cite{Kamra2019}, is provided by the intrinsically squeezed nature of the antiferromagnetic magnons~\cite{Kamra2016,Kamra2017,Kamra2019} (Fig. \ref{fig:sqmagnon}). Referring to a spin-flip residing on sublattice A (B) as a spin-up (-down) magnon, the antiferromagnetic eigenmodes are formed by two-mode squeezing~\cite{Gerry2004} between these spin-up and -down magnons~\cite{Kamra2016,Kamra2017}. Thus, a spin-up AFM squeezed-magnon is comprised of a coherent superposition of states with N+1 spin-up and N spin-down magnons, as depicted in Fig.\,\ref{fig:sqmagnon}, where N runs from zero to infinity~\cite{Kral1990,Nieto1997}. The average spin on each sublattice associated with one squeezed-magnon is thus much larger than its unit net spin. Any excitations, such as itinerant electrons, that exchange-couple to only one of the sublattices thus experience a much stronger interaction proportional to the average spin residing on the particular sublattice. The exposure of itinerant electrons to a fully uncompensated antiferromagnetic interface accomplishes this effect. 

The mechanism we propose in this paper appears to be mathematically analogous~\cite{Kamra2019,Leroux2018,Qin2018} to the one based on nonequilibrium squeezing of phonons~\cite{Knap2016,Babadi2017} proposed to explain the light-induced transient enhancement in $T_c$ observed experimentally~\cite{Fausti2011,Mankowsky2014,Mitrano2016}. Our mechanism, however, exploits the intrinsic, equilibrium squeezed nature of the antiferromagnetic magnons~\cite{Kamra2019} in contrast to the driven, transient squeezing achieved with phonons~\cite{Knap2016}. Moreover, our mechanism attributes the increase in $\lambda$ to an enhanced electron-magnon coupling~\cite{Kamra2019,Leroux2018} while Knap and coworkers find a renormalized, reduced electron hopping which alters the density of states to underpin a similar enhancement~\cite{Knap2016}.

\textit{Model}. -- We consider a bilayer consisting of a NM interfaced with an AFMI. The magnetic ground state is assumed to be an ordered staggered state where the staggered magnetization is taken to be along the $z$-direction. The essential physics does not depend on whether the $z$-direction is taken to be in or out of the interfacial plane. We take $\hbar = a = 1$, where $a$ is the lattice constant. The system is modeled by a Hamiltonian \, $H = H_{\text{AFMI}} + H_{\text{NM}} + H_{\text{int}}$,
with
\begin{align}
    &H_{\text{AFMI}} = J\!\sum_{\langle \bm{i}, \bm{j} \rangle} \bm{S}_{\bm{i}} \cdot \bm{S}_{\bm{j}}- K\sum_{\bm{i}}S^2_{\bm{i}z},
    \label{H_FM}\\
    &H_{\text{NM}} = -t\!\sum_{\langle \bm{i}, \bm{j} \rangle\sigma} c^{\dagger}_{\bm{i}\sigma}c_{\bm{j}\sigma} - \mu \sum_{\bm{i}\sigma}c^{\dagger}_{\bm{i}\sigma}c_{\bm{i}\sigma},\quad\quad\quad\quad\,\,\,\\
    &H_{\text{int}} = -2\Bar{J}_{A}\!\sum_{\bm{i}\in A}c_{\bm{i}}^{\dagger}\bm{\tau}c_{\bm{i}}\cdot \bm{S_i} -2\Bar{J}_{B}\!\sum_{\bm{i}\in B}c_{\bm{i}}^{\dagger}\bm{\tau}c_{\bm{i}}\cdot \bm{S_i}, 
    \label{H_int}
\end{align}
\noindent consisting of a term describing the AFMI, a tight binding Hamiltonian describing the NM, and a term accounting for exchange coupling between the two materials~\cite{Rohling2018,Takahashi2010,Zhang2012,Bender2015,Kamra2016,Ohnuma2013}. Here, $J~(>0)$ and $K~(>0)$ respectively parametrize the antiferromagnetic exchange and easy-axis anisotropy, and the sum over ${\langle \bm{i}, \bm{j} \rangle}$ includes all nearest neighbors for each $\bm{i}$. Further, $t$ is the tight-binding hopping parameter, the chemical potential is denoted by $\mu$ and $c_{i}^{\dagger} \equiv (c_{i\uparrow}^{\dagger}, c_{\bm{i}\downarrow}^{\dagger})$, where $c_{\bm{i}\sigma}^{\dagger}$ is a creation operator, creating an electron with spin $\sigma$ on lattice site $\bm{i}$. The Pauli matrices $\bm{\tau}$ act on the electron spin degree of freedom. Without loss of generality, the lattices on both sides of the interface are assumed to be square. The local exchange coupling between the NM electrons and AFMI spins across the interface is parametrized by sublattice-dependent strengths $\bar{J}_A$ and $\bar{J}_B$~\cite{Kamra2017B,Kamra2018}.

Employing Holstein-Primakoff transformations for the spin operators and switching to Fourier space, we obtain the magnetic Hamiltonian in terms of the sublattice magnons~\cite{SupplMat,Kittel1963}, which are not the eigenmodes. Executing Bogoliubov transformation brings the AFMI Hamiltonian to its diagonal form~\cite{SupplMat,Kittel1963}: $H_{\text{AFMI}}=\sum_{\bm{k}}\omega_{\bm{k}}\big(\alpha^{\dagger}_{\bm{k}}\alpha_{\bm{k}} + \beta^{\dagger}_{\bm{k}}\beta_{\bm{k}}\big)$, where $\omega_{\bm{k}} = 2s\sqrt{(zJ+K)^2-z^2J^2\gamma^2_{\bm{k}}}$, ${\gamma}_{\bm{k}} = 2\sum_{b}\cos(k_b)/z$, and the sum over $\bm{k}$ covers the reduced Brillouin zone of the sublattices. Here $z$ is the number of nearest neighbors, $s$ is the spin quantum number associated with the lattice site spins and the sum over $b$ covers the directions parallel to the interface. The magnon operators $\alpha_{k}$ and $\beta_{k}$ are coherent superpositions of the individual sublattice magnons $\alpha_{\bm{k}} = u_{\bm{k}}a_{\bm{k}} - v_{\bm{k}}b^{\dagger}_{-\bm{k}}$, $\beta_{\bm{k}} = u_{\bm{k}}b_{\bm{k}} - v_{\bm{k}}a^{\dagger}_{-\bm{k}}$, where $u_{\bm{k}} = \cosh(\theta_{\bm{k}})$, $v_{\bm{k}} = \sinh(\theta_{\bm{k}})$ and $\theta_{\bm{k}} = \frac{1}{2}\tanh^{-1}\Big(-\frac{Jz\gamma_{\bm{k}}}{zJ+K}\Big)$. Performing a Fourier transformation, the NM Hamiltonian becomes $H_{\text{NM}} = \sum_{\bm{k}\sigma}\epsilon_{\bm{k}}c^{\dagger}_{\bm{k}\sigma}c_{\bm{k}\sigma}$, where $\epsilon_{\bm{k}} = -tz\gamma_{\bm{k}} - \mu$. The sum over $\bm{k}$ here covers the full Brillouin zone.

As detailed in the supplemental material~\cite{SupplMat}(see, also, references \cite{Fjaerbu2017, Burdick1963, Lin2008, Ramchandani1970, Samuelsen1969, Kobler2006} therein), the interaction Hamiltonian [Eq. (\ref{H_int})] couples the NM electrons with the A and B sublattices of the AFMI:
\begin{align*}
    H^{(A)}_{\text{int}} &= \Omega V\sum_{\bm{k}\bm{q}}\Big[\Big(u_{\bm{q}}\alpha_{\bm{q}} + v_{\bm{q}}\beta^{\dagger}_{-\bm{q}}\Big)c^{\dagger}_{\bm{k}+\bm{q},\downarrow}c_{\bm{k}\uparrow} + \text{h.c.}\Big],\\
    H^{(B)}_{\text{int}} &= \,\,\,\,V\sum_{\bm{k}\bm{q}}\Big[\Big(u_{\bm{q}}\beta_{\bm{q}} + v_{\bm{q}}\alpha^{\dagger}_{-\bm{q}}\Big)c^{\dagger}_{\bm{k}+\bm{q},\uparrow}c_{\bm{k}\downarrow} + \text{h.c.}\Big],    
\end{align*}
where $V \equiv -\frac{2\Bar{J}\sqrt{s}}{\sqrt{N}}$, $N$ is the number of lattice sites in the interfacial plane and $\bar{J} = \bar{J}_B$.


\textit{Effective pairing interaction}. -- The full Hamiltonian now takes the form $ H = H_{\rm{AFMI}} + H_{\rm{NM}} + H^{(A)}_{\rm{int}} + H^{(B)}_{\rm{int}}$. In order to obtain an effective interacting theory for electrons in NM, we integrate out the magnons, using a canonical transformation \cite{Kittel1963}. We then obtain an effective electronic Hamiltonian $H'=H_{\rm{NM}} + H_{\rm{pair}}$~\cite{SupplMat}. The term $H_{\rm{pair}}$ contains the pairing interaction between electrons mediated by the antiferromagnetic magnons. Considering opposite momenta pairing~\cite{Schrieffer1999}, we obtain~\cite{SupplMat}
\begin{align}\label{eq:hpair}
    H_{\rm{pair}} &= \sum_{\bm{k}\bm{k}'}V_{\bm{k}\bm{k}'}c^{\dagger}_{\bm{k}\uparrow}c^{\dagger}_{-\bm{k}\downarrow}c_{-\bm{k}'\downarrow}c_{\bm{k}'\uparrow},
\end{align}
where 
\begin{align}
    V_{\bm{k}\bm{k}'} = - V^2 \frac{2\omega_{\bm{k} + \bm{k}'}}{(\epsilon_{\bm{k}'} - \epsilon_{\bm{k}} )^2 - \omega^2_{\bm{k} + \bm{k}'}}A(\bm{k} + \bm{k}', \Omega),
    \label{potential}
\end{align}
and
\begin{align}\label{eq:Afac}
    A(\bm{q}, \Omega) = \frac{1}{2}(\Omega^2 + 1)(u^2_{\bm{q}} + v^2_{\bm{q}}) + 2\,\Omega\, u_{\bm{q}} v_{\bm{q}}.
\end{align}
\noindent The interaction potential in Eq.\! \eqref{potential} consists of two factors, in addition to a prefactor. The first is the standard expression that enters pairing mediated by bosons with a dispersion relation $\omega_{\bm{k}}$, familiar from phonon-mediated superconductivity~\cite{Schrieffer1999}. The second factor $A(\bm{q}, \Omega)$ [Eq. \ref{eq:Afac}] contains the effect of the constructive or destructive interference of squeezed magnons. For long-wavelength magnons, the coherence factors $u_{\bm{q}}$ and $v_{\bm{q}}$ grow large, but have opposite signs. For the case of equal coupling to both sublattices, $\Omega = 1$, we have $A(\bm{q}, \Omega) = (u_{\bm{q}} + v_{\bm{q}})^2$, and a near-cancellation of the coherence factors. On the other hand, for the case of a fully uncompensated AFMI interface, $\Omega = 0$, the coherence factors combine to $A({\bm{q}},\Omega)=(u_{\bm{q}}^2+v_{\bm{q}}^2)/2$, where $u_{\bm{q}}$ and $v_{\bm{q}}$ are squared separately. This represents a dramatic amplification of the pairing interaction in the latter case. 


\textit{Mean field theory and $T_c$}. -- We now formulate the weak-coupling mean field theory for the magnon-mediated superconductivity in the NM employing standard methodology for unconventional superconductors~\cite{Sigrist1991}. Comparing our effective interaction potential [Eqs. (\ref{eq:hpair}) and (\ref{potential})] with that for conventional s-wave superconductors~\cite{Schrieffer1999}, we note the additional multiplicative minus sign. This implies that the conventional spin-singlet pairing channel is repulsive and does not support condensation. We therefore consider $S_z=0$ spin-triplet pairing which is the typical condensation channel for magnon-mediated superconductivity~\cite{Karchev2003,Rohling2018,Pfleiderer2009,Mackenzie2003}. The corresponding gap function is defined as $\Delta_{\bm{k}} = -\sum_{\bm{k}'}V_{\bm{k}\bm{k}',O(\bm{k})}\langle c_{-\bm{k}'\uparrow}c_{\bm{k}'\downarrow} + c_{-\bm{k}'\downarrow}c_{\bm{k}'\uparrow}\rangle/2$, where $V_{\bm{k}\bm{k}',O(\bm{k})} = \frac{1}{2}(V_{\bm{k}\bm{k}'} - V_{-\bm{k},\bm{k}'})$ is the odd part of the pairing potential [Eq. (\ref{potential})]. The ensuing gap equation takes the form~\cite{Sigrist1991}
\begin{align}
    \Delta_{\bm{k}} = -\sum_{\bm{k}'}V_{\bm{k}\bm{k}',O(\bm{k})}\frac{\Delta_{\bm{k}'}}{2E_{\bm{k}'}}\tanh(\frac{E_{\bm{k}'}}{2k_B T}),
\end{align}
\noindent where $E_{\bm{k}} = \sqrt{\epsilon^2_k + |\Delta_{\bm{k}}|^2}$. Close to the critical temperature, we linearize the gap equation and compute an average over the Fermi surface $\lambda \Delta_{\bm{k}} = - D_{0} \langle V_{\bm{k}\bm{k}',O(\bm{k})} \Delta_{\bm{k}'} \rangle_{\bm{k}', \text{FS}}$, in order to determine the critical temperature~\cite{Sigrist1991,SupplMat}
\begin{align}
    k_B T_c = 1.14\,\omega_c\,e^{-1/\lambda}.   
\end{align}
Here, $D_0$ is the density of states on the Fermi surface and $\omega_c$ is the high-frequency cutoff on the magnon spectrum. As detailed in the supplemental material~\cite{SupplMat}, numerical solution of the eigenvalue problem for the dimensionless coupling constant $\lambda$ yields a p-wave gap function. Employing experimentally obtained material parameters~\cite{SupplMat}, critical temperature $T_c$ and dimensionless coupling constant $\lambda$ are evaluated and presented as a function of the asymmetry parameter $\Omega \equiv \bar{J}_A/\bar{J}_B$ in Fig.~\ref{fig:Tc}. We now pause to comment on the results thus obtained.

Both $\lambda$ and $T_c$ are found to increase with the sublattice-asymmetry of the interfacial exchange coupling, i.e. as $\Omega$ decreases (Fig.~\ref{fig:Tc}). Furthermore, a $T_c$ of few ten Kelvins is obtained for a perfectly uncompensated interface, corresponding to $\Omega = 0$, employing realistic parameters~\cite{SupplMat}. However, $T_c$ evaluations are notoriously unreliable on account of their sensitivity to $\lambda$ (exponential dependence on $1/\lambda$) evaluation method and related details. Within our model, altering the parameters by a few ten percents leads to significantly smaller, or even larger, critical temperatures. The key lesson of our analysis is that uncompensated interfaces drastically enhance $T_c$ to values potentially significantly larger than 1 K considering realistic materials.  

\textit{Theoretical model assessment}. -- Uncompensated AFM interfaces induce spin splitting in the adjacent conductor~\cite{Kamra2018} that has not been included here. In conventional BCS superconductors this effect has been investigated in detail and is known to result in rich physics~\cite{Bergeret2005,Bergeret2018} including gapless superconductivity~\cite{Maki1969}, FFLO state~\cite{Fulde1964,Larkin1965}, and finally, destruction of the superconducting phase when spin splitting becomes significantly larger than $T_c$~\cite{Maki1964}. In the present case, the magnon-squeezing effect amplifies the electron-magnon coupling, and thus $T_c$, while leaving the spin splitting unchanged. Thus, the latter is expected to bear no significant effects on the predicted superconducting state~\cite{Mackenzie2003} with $T_c$ considerably larger than the typically induced spin splitting $\sim 1$~K~\cite{Li2013,Kamra2018}. Spin-splitting may also be suppressed by applying an external compensating magnetic field \cite{PhysRevLett.90.197006}.  The system considered in this paper
is far less susceptible to non-trivial feedback effects of the itinerant electrons on the antiferromagnet than the case studied in Ref.\! \onlinecite{PhysRevLett.119.247203}, particularly since the magnetic surface we consider is the surface of a bulk magnet and an Ising easy-axis anisotropy is included in the description.  A strong spin-orbit interaction, also disregarded here, may reduce $T_c$~\cite{Bergeret2018}. Finally, all non s-wave superconducting phases are suppressed by disorder. Therefore, we expect the inclusion of interfacial disorder to reduce $T_c$ for our p-wave state~\cite{Mackenzie2003}. A rigorous analysis of the effects mentioned above constitutes a promising avenue for future studies.

\textit{Experimental feasibility}. -- The fabrication of proposed bilayers with uncompensated and low disorder interfaces, albeit challenging, is within the reach of contemporary state-of-the-art techniques~\cite{Zhang2016,Gong2017}. The choice of materials is likely to be dictated by the growth, rather than theoretical, considerations. Nevertheless, we now outline the optimal materials requirements from a theory standpoint. Broadly speaking, a reasonably large N{\'e}el temperature for the AFM is beneficial. A metal with high density of states at the Fermi level and low spin-orbit interaction is desirable. The possibility of a strong exchange coupling across the interface seems to be supported by spin mixing conductance experiments for a wide range of bilayers~\cite{Czeschka2011,Weiler2013,Kajiwara2010}. Without an extensive comparison between many materials, hematite~\cite{Lebrun2018} or chromia~\cite{He2010,Kosub2017} as AFM and copper or aluminum as the metal seem to be reasonable choices.

\textit{Summary}. -- We have shown that magnons in an antiferromagnetic insulator mediate attractive electron-electron interactions in an adjacent normal metal. Exploiting the intrinsic squeezing of antiferromagnetic magnons, the electron-electron pairing potential is amplified by exchange coupling the normal metal asymmetrically to the two sublattices of the antiferromagnet. This, in turn, is found to result in a dramatic increase in the superconducting critical temperature, which is estimated to be significantly larger than 1 K employing experimentally obtained material parameters, when the normal metal is exposed to an uncompensated antiferromagnetic interface. Our results demonstrate the possibility of engineering heterostructures exhibiting superconductivity at potentially large temperatures.

We thank Jagadeesh Moodera, Rudolf Gross, Stephan Gepr{\"a}gs, Matthias Althammer, Niklas Rohling and Michael Knap for valuable discussions. We acknowledge financial support from the Research Council of Norway Grant No. 262633 ``Center of Excellence on Quantum Spintronics'', and Grant No. 250985, ``Fundamentals of Low-dissipative Topological Matter''.

\bibliography{Refs}

\clearpage


\onecolumngrid
\setlength\parindent{0pt}

\begin{center}
  \textbf{\large Supplementary material\\ Enhancement of superconductivity mediated by antiferromagnetic squeezed magnons}\\[.4cm]
  Eirik Erlandsen, Akashdeep Kamra, Arne Brataas, and Asle Sudbø$^*$\\
  {\itshape Center for Quantum Spintronics, Department of Physics, Norwegian University of Science and Technology,\\ NO-7491 Trondheim, Norway}\\[1.0cm]
\end{center}


In this supplement, we provide more details for the derivations of the results presented in the main paper. In the following we will take $\hbar = a = 1$.

\section{Model}

\begin{figure}[h] 
    \begin{center}
        \includegraphics[width=0.28\columnwidth,trim= 6.0cm 10.5cm 6.0cm 11.0cm,clip=true]{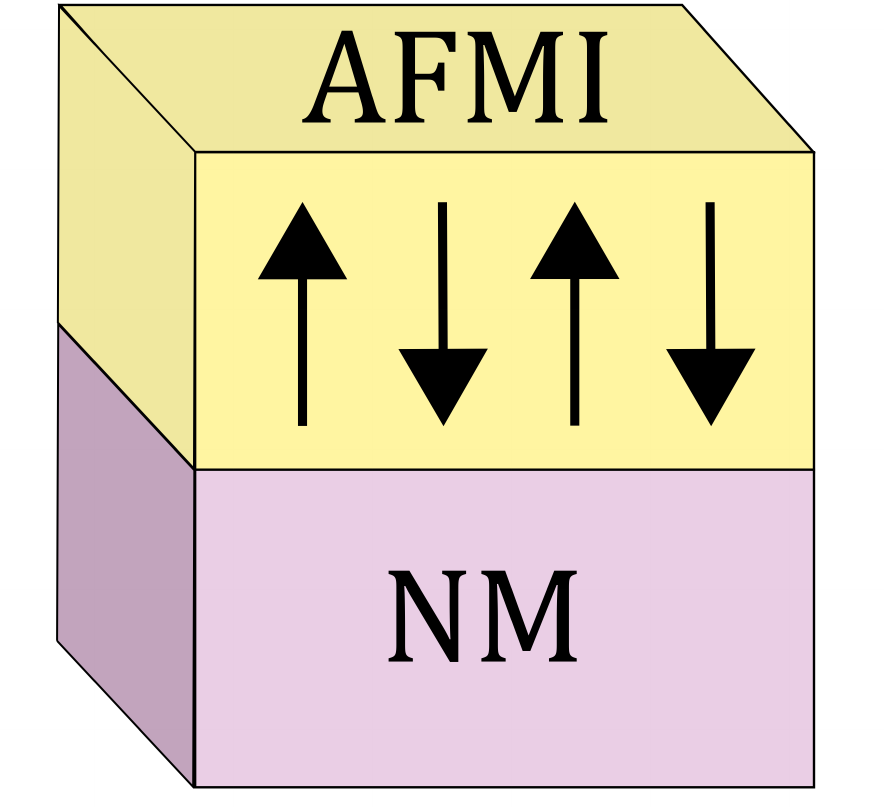}
    \end{center}
    \caption{The system consists of an antiferromagnetic insulator (AFMI) placed on top of a normal metal (NM).}
    \label{fig:NM_Boxes}
\end{figure}

We consider a bilayer heterostructure consisting of a normal metal (NM) and an antiferromagnetic insulator (AFMI), as shown in Fig.\! \ref{fig:NM_Boxes}. The staggered magnetization of the AFMI is assumed to be aligned with the $z$-direction, which could be either in-plane or out-of-plane.  

The system is modeled by the  Hamiltonian $H = H_{\text{AFMI}} + H_{\text{NM}} + H_{\text{int}}$ \cite{Fjaerbu2018, Fjaerbu2019}, where 

\begin{align}
    H_{\text{AFMI}} &= J\!\sum_{\langle \bm{i}, \bm{j} \rangle} \bm{S}_{\bm{i}} \cdot \bm{S}_{\bm{j}}- K\sum_{\bm{i}}S^2_{\bm{i}z},\\
    H_{\text{NM}}\,\,\,\,\, &=  -t\!\sum_{\langle \bm{i}, \bm{j} \rangle\sigma} c^{\dagger}_{\bm{i}\sigma}c_{\bm{j}\sigma} - \mu \sum_{\bm{i}\sigma}c^{\dagger}_{\bm{i}\sigma}c_{\bm{i}\sigma},\\
    H_{\text{int}}\,\,\,\,\,\,\, &= -2\Bar{J}_{A}\sum_{\bm{i}\in A}c_{\bm{i}}^{\dagger}\bm{\tau}c_{\bm{i}}\cdot \bm{S_i} -2\Bar{J}_{B}\sum_{\bm{i}\in B}c_{\bm{i}}^{\dagger}\bm{\tau}c_{\bm{i}}\cdot \bm{S_i}. 
\end{align}

Here, we have $c_{i}^{\dagger} = (c_{i\uparrow}^{\dagger}, c_{i\downarrow}^{\dagger})$, where $c_{i\sigma}^{\dagger}$ is a creation operator, creating an electron with spin $\sigma$ on lattice site $\bm{i}$ in the NM. The chemical potential is denoted by $\mu$. The exchange coefficients $J$ is assumed to be positive and therefore favors anti-alignment of neighboring lattice site spins $\bm{S}_{\bm{i}}$. The easy-axis anisotropy coefficient $K$ is also positive. The Pauli matrices $\bm{\tau}$ act on the fermionic spin degree of freedom, the lattices are taken to be square and we have periodic boundary conditions in the directions parallel to the interfacial plane. The sum over ${\langle \bm{i}, \bm{j} \rangle}$ includes all nearest neighbors for each $\bm{i}$, and the lattice site sums in the interaction Hamiltonian cover the interfacial plane between the two materials. The strength of the coupling between the electrons and the lattice site spins of sublattice $A$, $B$ is determined by the parameters $\Bar{J}_{A}$, $\Bar{J}_{B}$. In the following, we will take $\Bar{J}_{B} = \Bar{J}$ and $\Bar{J}_{A} = \Omega \Bar{J}$, where $\Omega$ determines which AFMI sublattice couples strongest to the electrons on the surface of the NM. In this way of parametrizing the exchange-interaction across the AFMI-NM interface, we may without loss of generality set $0 \leq \Omega \leq 1$. \\  

We introduce Holstein-Primakoff transformations for both sublattices in the AFMI

\begin{align}
    S_{\bm{i}+}^A &= \sqrt{2s - a_{\bm{i}}^{\dagger}a_{\bm{i}}}\,\,a_{\bm{i}} \approx \sqrt{2s}\,a_{\bm{i}},\\
    S_{\bm{i}-}^A &= a_{\bm{i}}^{\dagger}\sqrt{2s-a_{\bm{i}}^{\dagger}a_{\bm{i}}} \,\approx \sqrt{2s}\,a_{\bm{i}}^{\dagger},\\
    S_{\bm{i}z}^A\, &= s - a_{\bm{i}}^{\dagger}a_{\bm{i}},
\end{align}

\begin{align}
    S_{\bm{j}+}^B &= b_{\bm{j}}^{\dagger}\sqrt{2s - b_{\bm{j}}^{\dagger}b_{\bm{j}}} \,\approx \sqrt{2s}\,b_{\bm{j}}^{\dagger},\\
    S_{\bm{j}-}^B &= \sqrt{2s-b_{\bm{j}}^{\dagger}b_{\bm{j}}}\,\,b_{\bm{j}} \approx \sqrt{2s}\,b_{\bm{j}},\\
    S_{\bm{j}z}^B\, &= -s + b_{\bm{j}}^{\dagger}b_{\bm{j}},
\end{align}

and Fourier transformations for the magnon and electron operators

\begin{align}
    a_{\bm{i}} = \frac{1}{\sqrt{N_A}}\sum_{\bm{k}\in\diamondsuit}a_{\bm{k}}e^{-i\bm{k}\cdot\bm{r}_{\bm{i}}}, \quad\quad b_{\bm{i}} = \frac{1}{\sqrt{N_B}}\sum_{\bm{k}\in\diamondsuit}b_{\bm{k}}e^{-i\bm{k}\cdot\bm{r}_{\bm{i}}},
\end{align}

\begin{align}
    c_{\bm{i}\sigma} &= \frac{1}{\sqrt{N}} \sum_{\bm{k}\in \diamondsuit}\Big(c_{\bm{k}\sigma}e^{-i\bm{k}\cdot\bm{r}_{\bm{i}}} + c_{\bm{k}+\bm{G},\sigma}e^{-i(\bm{k}+ \bm{G})\cdot\bm{r}_{\bm{i}}}\Big),
\end{align}

where $\diamondsuit$ indicates that the sum over momenta covers the reduced Brillouin zone of the sublattices and $\bm{G} \equiv \frac{\pi(\hat{x} + \hat{y})}{a}$ is a reciprocal lattice vector for the sublattices. After performing a Bogoliubov transformation, the AFMI Hamiltonian becomes

\begin{align}
    H_{\text{AFMI}} = \sum_{\bm{k}\in\diamondsuit}\omega_{\bm{k}}\Big(\alpha^{\dagger}_{\bm{k}}\alpha_{\bm{k}} + \beta^{\dagger}_{\bm{k}}\beta_{\bm{k}}\Big), 
\end{align}

with

\begin{align}
    \omega_{\bm{k}} = 2s\sqrt{(zJ+K)^2-z^2J^2\gamma^2_{\bm{k}}},
\end{align}

\begin{align}
    {\gamma}_{\bm{k}} = \frac{2}{z}\sum_{b}\cos(k_b),  
    \label{sum_b}
\end{align}

\begin{align}
    \alpha_{\bm{k}} = u_{\bm{k}}a_{\bm{k}} - v_{\bm{k}}b^{\dagger}_{-\bm{k}}, \quad\quad \beta_{\bm{k}} = u_{\bm{k}}b_{\bm{k}} - v_{\bm{k}}a^{\dagger}_{-\bm{k}},
\end{align}

\begin{align}
    u_{\bm{k}} = \cosh(\theta_{\bm{k}}), \quad\quad v_{\bm{k}} = \sinh(\theta_{\bm{k}}),
\end{align}

\begin{align}
    \theta_{\bm{k}} = \frac{1}{2}\tanh^{-1}\Big(-\frac{Jz\gamma_{\bm{k}}}{zJ+K}\Big).
\end{align}

The number of nearest neighbors is here denoted by $z$, and the sum over $b$ in Eq.\! \eqref{sum_b} goes over the directions parallel to the interface. For small $k$, compared to the size of the Brillouin zone, the parameters $u_{\bm{k}}$ and $v_{\bm{k}}$ grow large with similar magnitude, but opposite signs.\\

The NM Hamiltonian reduces to 

\begin{align}
    H_{\text{NM}} = \sum_{\substack{\bm{k}\in\Box \\ \sigma}}\epsilon_{\bm{k}}c^{\dagger}_{\bm{k}\sigma}c_{\bm{k}\sigma},
\end{align}

with

\begin{align}
    \epsilon_{\bm{k}} = -tz\gamma_{\bm{k}} - \mu.
\end{align}

From the interaction Hamiltonian we obtain, for coupling to sublattice $A$ and $B$ respectively \cite{Fjaerbu2017, Fjaerbu2019},

\begin{align}
    H^{(A)}_{\text{int}} = \Omega V\sum_{\substack{\bm{k} \in \Box\\ \bm{q} \in \diamondsuit}}\Bigg(a_{\bm{q}}c^{\dagger}_{\bm{k} + \bm{q},\downarrow}c_{\bm{k}\uparrow} + a_{\bm{q}}c^{\dagger}_{\bm{k} + \bm{q} +\bm{G},\downarrow}c_{\bm{k}\uparrow} + \text{h.c.}\Bigg) - \Omega\Bar{J}s\sum_{\substack{\bm{k}\in\Box\\ \sigma}}\hat{\sigma}\Bigg(c^{\dagger}_{\bm{k}\sigma}c_{\bm{k}\sigma} + c^{\dagger}_{\bm{k}+\bm{G},\sigma}c_{\bm{k}\sigma}\Bigg),
\end{align}

\begin{align}
    H^{(B)}_{\text{int}} = V\sum_{\substack{\bm{k} \in \Box\\ \bm{q} \in \diamondsuit}}\Bigg(b_{\bm{q}}c^{\dagger}_{\bm{k} + \bm{q},\uparrow}c_{\bm{k}\downarrow} - b_{\bm{q}}c^{\dagger}_{\bm{k} + \bm{q} +\bm{G},\uparrow}c_{\bm{k}\downarrow} + \text{h.c.}\Bigg) + \Bar{J}s\sum_{\substack{\bm{k}\in\Box\\ \sigma}}\hat{\sigma}\Bigg(c^{\dagger}_{\bm{k}\sigma}c_{\bm{k}\sigma} - c^{\dagger}_{\bm{k}+\bm{G},\sigma}c_{\bm{k}\sigma}\Bigg),\,\,\,\,\,\,\,\,\,\,\,\,\,
\end{align}

where $\hat{\sigma} = \pm 1$ depending on the spin being up or down. We have also defined 

\begin{align}
    V \equiv -\frac{2\Bar{J}\sqrt{s}}{\sqrt{N}},
\end{align}

where $N$ is the number of lattice sites in the interfacial plane and used $\Box$ to mark the sums that cover the Brillouin zone of the full lattice. Quadratic or higher order terms in the magnon operators have been neglected. The relative minus signs between the two terms in each of the parentheses in the expression for $H^{(B)}_{\text{int}}$ arise due to sublattice $B$ being shifted in space one lattice constant away from sublattice $A$.\\

For our tight binding NM model, the different sides of the Fermi surface are connected by a reciprocal lattice vector $\bm{G}$, in the case of half-filling. The above Umklapp processes involving $\bm{G}$ are then important for the physics at the Fermi surface \cite{Fjaerbu2019}. In the following, we focus on the case away from half-filling and neglect such processes. Moreover, based on experimental results, the effect of the potential Zeeman splitting is expected to be small and a rigorous treatment of the corrections to the superconducting state stemming from this effect is outside the scope of this letter. See the discussion in the main paper. We therefore neglect these terms as well and obtain 

\begin{align}
    H_{\text{int}} = V\sum_{\substack{\bm{k} \in \Box\\ \bm{q} \in \diamondsuit}}\Big(\Omega\,a_{\bm{q}}c^{\dagger}_{\bm{k}+\bm{q},\downarrow}c_{\bm{k}\uparrow} + b_{\bm{q}}c^{\dagger}_{\bm{k}+\bm{q},\uparrow}c_{\bm{k}\downarrow} + \text{h.c.}\Big).
\end{align}

Rewriting the magnon operators in terms of the quasiparticles that diagonalized the AFMI Hamiltonian, we then have

\begin{align}
    H_{\text{int}} = V\sum_{\bm{k}\bm{q}}\Bigg[\Omega\Big(u_{\bm{q}}\alpha_{\bm{q}} + v_{\bm{q}}\beta^{\dagger}_{-\bm{q}}\Big)c^{\dagger}_{\bm{k}+\bm{q},\downarrow}c_{\bm{k}\uparrow} + \Big(u_{\bm{q}}\beta_{\bm{q}} + v_{\bm{q}}\alpha^{\dagger}_{-\bm{q}}\Big)c^{\dagger}_{\bm{k}+\bm{q},\uparrow}c_{\bm{k}\downarrow} + \text{h.c.}\Bigg].
\end{align}

Collecting the results, the total Hamiltonian takes the form $H = H_{\text{AFMI}} + H_{\text{NM}} + H^{(A)}_{\text{int}} + H^{(B)}_{\text{int}}$,

\begin{align}
    H_{\text{AFMI}} &= \sum_{\bm{k}}\omega_{\bm{k}}\Big(\alpha^{\dagger}_{\bm{k}}\alpha_{\bm{k}} + \beta^{\dagger}_{\bm{k}}\beta_{\bm{k}}\Big),\\
    H_{\text{NM}}\,\,\,\,\, &=  \sum_{\bm{k}\sigma}\epsilon_{\bm{k}}c^{\dagger}_{\bm{k}\sigma}c_{\bm{k}\sigma},\\
    H^{(A)}_{\text{int}}\,\,\,\,\, &= \Omega V\sum_{\bm{k}\bm{q}}\Bigg[\Big(u_{\bm{q}}\alpha_{\bm{q}} + v_{\bm{q}}\beta^{\dagger}_{-\bm{q}}\Big)c^{\dagger}_{\bm{k}+\bm{q},\downarrow}c_{\bm{k}\uparrow} + \text{h.c.}\Bigg],\\
    H^{(B)}_{\text{int}}\,\,\,\,\, &= \,\,\,\,V\sum_{\bm{k}\bm{q}}\Bigg[\Big(u_{\bm{q}}\beta_{\bm{q}} + v_{\bm{q}}\alpha^{\dagger}_{-\bm{q}}\Big)c^{\dagger}_{\bm{k}+\bm{q},\uparrow}c_{\bm{k}\downarrow} + \text{h.c.}\Bigg].
\end{align}

\section{Effective Interaction}

We now perform a canonical transformation in order to eliminate the magnon operators from the problem and obtain an effective interacting  theory for the electrons, with the electron-electron interaction mediated by virtual magnons. We define

\begin{align}
    &H_0 \equiv \sum_{\bm{k}}\omega_{\bm{k}}\Big(\alpha^{\dagger}_{\bm{k}}\alpha_{\bm{k}} + \beta^{\dagger}_{\bm{k}}\beta_{\bm{k}}\Big) + \sum_{\bm{k}\sigma}\epsilon_{\bm{k}\sigma}c^{\dagger}_{\bm{k}\sigma}c_{\bm{k}\sigma}\\
    &\eta H_1 = \eta H_1^{(A)} + \eta H_1^{(B)} \equiv H_{\text{int}}^{(A)} + H_{\text{int}}^{(B)},
\end{align}

and write

\begin{align}
\begin{aligned}
    H'' &= e^{-\eta S}H\,e^{\eta S} = H + \eta\comm{H}{S} + \frac{\eta^2}{2!}\comm{\comm{H}{S}}{S} + \mathcal{O}(\eta^3)\\
    &= H_0 + \eta\Big(H_1 + \comm{H_0}{S}\Big) + \eta^2\Big(\comm{H_1}{S} + \frac{1}{2}\comm{\comm{H_0}{S}}{S} \Big) + \mathcal{O}(\eta^3).
\end{aligned}
\end{align}

We then choose $\eta S = \eta S^{(A)} + \eta S^{(B)}$ such that we have

\begin{align}
    \eta H_1^{(L)} + \comm{H_0}{\eta S^{(L)}} = 0,
\end{align}

producing 

\begin{align}
    H' = H_0 + \frac{1}{2}\sum_{LL'}\comm{\eta H_1^{(L)}}{\eta S^{(L')}} + \mathcal{O}(\eta^3),
\end{align}

where $L\in \{A,B\}$. Choosing 

\begin{align}
    \eta S^{(A)} &= \Omega V\sum_{\bm{k}\bm{q}}\Bigg[\Big(x_{\bm{k},\bm{q}}u_{\bm{q}}\alpha_{\bm{q}} + y_{\bm{k},\bm{q}}v_{\bm{q}}\beta^{\dagger}_{-\bm{q}}\Big)c^{\dagger}_{\bm{k}+\bm{q},\downarrow}c_{\bm{k}\uparrow} + \Big(z_{\bm{k},\bm{q}}u_{\bm{q}}\alpha^{\dagger}_{-\bm{q}} + w_{\bm{k},\bm{q}}v_{\bm{q}}\beta_{\bm{q}}\Big)c^{\dagger}_{\bm{k}+\bm{q},\uparrow}c_{\bm{k}\downarrow}\Bigg],\\
    \eta S^{(B)} &= \,\,\,\,\, V\sum_{\bm{k}\bm{q}}\Bigg[\Big(w_{\bm{k},\bm{q}}u_{\bm{q}}\beta_{\bm{q}} + z_{\bm{k},\bm{q}}v_{\bm{q}}\alpha^{\dagger}_{-\bm{q}}\Big)c^{\dagger}_{\bm{k}+\bm{q},\uparrow}c_{\bm{k}\downarrow} + \Big(y_{\bm{k},\bm{q}}u_{\bm{q}}\beta^{\dagger}_{-\bm{q}} + x_{\bm{k},\bm{q}}v_{\bm{q}}\alpha_{\bm{q}}\Big)c^{\dagger}_{\bm{k}+\bm{q},\downarrow}c_{\bm{k}\uparrow}\Bigg],
\end{align}

where

\begin{align}
    x_{\bm{k},\bm{q}} &= w_{\bm{k},\bm{q}} = \frac{1}{\epsilon_{\bm{k}} - \epsilon_{\bm{k}+\bm{q}} + \omega_{\bm{q}}},\quad\quad y_{\bm{k},\bm{q}} = z_{\bm{k},\bm{q}} = \frac{1}{\epsilon_{\bm{k}} - \epsilon_{\bm{k} + \bm{q}} - \omega_{\bm{q}}},
\end{align}

and working out the commutators, one obtains

\begin{align}
\begin{aligned}
    H^{(A,A)}_{\text{pair}} &= \frac{1}{2}\Omega^2V^2\sum_{\bm{k}\bm{q}\bm{k}'}\,c^{\dagger}_{\bm{k}+\bm{q}\downarrow}c_{\bm{k}\uparrow}c^{\dagger}_{\bm{k}'-\bm{q}\uparrow}c_{\bm{k}'\downarrow}
    \Bigg[u^2_{\bm{q}}\Bigg(\frac{1}{(\epsilon_{\bm{k}'} - \epsilon_{\bm{k}'-\bm{q}}) - \omega_{\bm{q}}} - \frac{1}{(\epsilon_{\bm{k}} - \epsilon_{\bm{k} + \bm{q}}) + \omega_{\bm{q}}}\Bigg)\\
    &\quad\quad\quad\quad\quad\quad\quad\quad+  v^2_{\bm{q}}\Bigg(\frac{1}{(\epsilon_{\bm{k}} - \epsilon_{\bm{k}+\bm{q}}) -\omega_{\bm{q}}} - \frac{1}{(\epsilon_{\bm{k}'} - \epsilon_{\bm{k}' - \bm{q}}) +\omega_{\bm{q}}}\Bigg)\Bigg],
\end{aligned}
\end{align}

\begin{align}
\begin{aligned}
    H^{(B,B)}_{\text{pair}} &= \frac{1}{2}V^2\sum_{\bm{k}\bm{q}\bm{k}'}\,c^{\dagger}_{\bm{k}+\bm{q}\downarrow}c_{\bm{k}\uparrow}c^{\dagger}_{\bm{k}'-\bm{q}\uparrow}c_{\bm{k}'\downarrow}
    \Bigg[v^2_{\bm{q}}\Bigg(\frac{1}{(\epsilon_{\bm{k}'} - \epsilon_{\bm{k}'-\bm{q}}) -\omega_{\bm{q}}} - \frac{1}{(\epsilon_{\bm{k}} - \epsilon_{\bm{k} + \bm{q}}) +\omega_{\bm{q}}}\Bigg)\\
    &\quad\quad\quad\quad\quad\quad\quad\quad+  u^2_{\bm{q}}\Bigg(\frac{1}{(\epsilon_{\bm{k}} - \epsilon_{\bm{k}+\bm{q}}) -\omega_{\bm{q}}} - \frac{1}{(\epsilon_{\bm{k}'} - \epsilon_{\bm{k}' - \bm{q}}) +\omega_{\bm{q}}}\Bigg)\Bigg],
\end{aligned}
\end{align}

\begin{align}
\begin{aligned}
    &H^{(A,B)}_{\text{pair}} + H^{(B,A)}_{\text{pair}} = \Omega V^2\sum_{\bm{k}\bm{q}\bm{k}'}\,c^{\dagger}_{\bm{k}+\bm{q}\downarrow}c_{\bm{k}\uparrow}c^{\dagger}_{\bm{k}'-\bm{q}\uparrow}c_{\bm{k}'\downarrow}\,u_{\bm{q}}v_{\bm{q}}\Bigg(\frac{1}{(\epsilon_{\bm{k}'} - \epsilon_{\bm{k}' - \bm{q}}) - \omega_{\bm{q}}}\\
    &+ \frac{1}{(\epsilon_{\bm{k}} - \epsilon_{\bm{k}+\bm{q}}) -\omega_{\bm{q}}} - \frac{1}{(\epsilon_{\bm{k}'} - \epsilon_{\bm{k}' - \bm{q}}) + \omega_{\bm{q}}} - \frac{1}{(\epsilon_{\bm{k}} - \epsilon_{\bm{k}+\bm{q}}) +\omega_{\bm{q}}}\Bigg).
\end{aligned}
\end{align}

Here, we have defined $H^{(L,L')}_{\text{pair}}$ as the contribution from $\frac{1}{2}\comm{\eta H_1^{(L)}}{\eta S^{(L')}}$ that takes the form of an electron-electron interaction. Collecting together the different contributions and focusing on BCS-type pairing between electrons on opposite sides of the Fermi surface, the result can be written on the following form

\begin{align}
    H_{\rm{pair}} &= \sum_{\bm{k}\bm{k}'}V_{\bm{k}\bm{k}'}c^{\dagger}_{\bm{k}\uparrow}c^{\dagger}_{-\bm{k}\downarrow}c_{-\bm{k}'\downarrow}c_{\bm{k}'\uparrow},
\end{align}

where 

\begin{align}
    V_{\bm{k}\bm{k}'} = - V^2 \frac{2\omega_{\bm{k} + \bm{k}'}}{(\epsilon_{\bm{k}'} - \epsilon_{\bm{k}} )^2 - \omega^2_{\bm{k} + \bm{k}'}}A(\bm{k} + \bm{k}', \Omega),
    \label{potential_sup}
\end{align}

and

\begin{align}
    A(\bm{k} + \bm{k}', \Omega) = \frac{1}{2}(\Omega^2 + 1)(u^2_{\bm{k} + \bm{k}'} + v^2_{\bm{k} + \bm{k}'}) + 2\,\Omega\, u_{\bm{k} + \bm{k}'} v_{\bm{k} + \bm{k}'}.
\end{align}

The fraction in Eq.\! \eqref{potential_sup} is of the standard form for electron-electron interactions mediated by a boson. The $A$-factor quantifies the effect of the interference between squeezed magnon states \cite{Kamra2016,Kamra2017} on sublattices $A$ and $B$. Assuming $q$ significantly smaller than the size of the Brillouin zone, the term involving $u^2_{\bm{q}} + v^2_{\bm{q}}$ grows large and positive, while the next term involving $u_{\bm{q}}v_{\bm{q}}$ grows large and negative, due to the opposite signs of the parameters $u_{\bm{q}}$ and $v_{\bm{q}}$. The destructive interference between squeezed magnon states is in general maximal when $\Omega = 1$. Then, the factor within the square brackets simplify to $(u_{\bm{q}} + v_{\bm{q}})^2$, which for general filling fractions is small due to a near cancellation of $u_{\bm{q}}$ and $v_{\bm{q}}$. Setting instead $\Omega=0$ eliminates the destructive interference between squeezed magnon states on sublattices $A$ and $B$ entirely.\\

\section{Gap equation}

The potential can be divided into even and odd parts in $\bm{k}$ 

\begin{align}
    V_{\bm{k}\bm{k}'} = V_{\bm{k}\bm{k}',O(\bm{k})} + V_{\bm{k}\bm{k}',E(\bm{k})},  
\end{align}

where

\begin{align}
    &V_{\bm{k}\bm{k}',O(\bm{k})} = \frac{1}{2}(V_{\bm{k}\bm{k}'} - V_{-\bm{k},\bm{k}'}),\\
    &V_{\bm{k}\bm{k}',E(\bm{k})} = \frac{1}{2}(V_{\bm{k}\bm{k}'} + V_{-\bm{k},\bm{k}'}).
\end{align}

Following the procedure of Ref.\! \cite{Sigrist1991}, we write the pairing Hamiltonian as 

\begin{align}
    H^{\rm{(BCS)}}_{\text{pair}} = \frac{1}{2}\sum_{\bm{k}\bm{k}'}\sum_{s_1 s_2 s_3 s_4}V^{s_1 s_2 s_3 s_4}_{\bm{k}\bm{k}'}c^{\dagger}_{\bm{k}s_1}c^{\dagger}_{-\bm{k} s_2}c_{-\bm{k}'s_3}c_{\bm{k}'s_4},
\end{align}

where 

\begin{align}
    &V^{\uparrow \downarrow \downarrow \uparrow}_{\bm{k}\bm{k}'} = V^{\downarrow \uparrow \uparrow \downarrow}_{\bm{k}\bm{k}'} = \frac{1}{2}\big[V_{\bm{k}\bm{k}',O(\bm{k})} + V_{\bm{k}\bm{k}',E(\bm{k})}\big]\\
    &V^{\downarrow \uparrow \downarrow \uparrow}_{\bm{k}\bm{k}'} =  V^{\uparrow \downarrow \uparrow \downarrow}_{\bm{k}\bm{k}'} = \frac{1}{2}\big[V_{\bm{k}\bm{k}',O(\bm{k})} - V_{\bm{k}\bm{k}',E(\bm{k})}\big],
\end{align}

following from the fermionic anti-commutation relations of the electron operators. The potential vanishes for all other spin-combinations. We define a gap function

\begin{align}
    \Delta_{\bm{k},s_1 s_2} = - \sum_{\bm{k}',s_3 s_4} V^{s_1 s_2 s_3 s_4}_{\bm{k}\bm{k}'}b_{\bm{k}', s_3 s_4},
\end{align}

where $b_{\bm{k},ss'} = \langle c_{-\bm{k}s}c_{\bm{k}s'}\rangle$. Following the usual mean-field approach, the gap equation then becomes

\begin{align}
    \Delta_{\bm{k},s_1 s_2} = - \sum_{\bm{k}', s_3 s_4}V^{s_1 s_2 s_3 s_4}_{\bm{k}\bm{k}'}\Delta_{\bm{k}',s_4 s_3}\chi_{\bm{k}'},
\end{align}

where

\begin{align}
    \chi_{\bm{k}'} = \frac{1}{2E_{\bm{k}'}}\tanh(\frac{E_{\bm{k}'}}{2k_B T}),
\end{align}

and $E_{\bm{k}} = \sqrt{\epsilon^2_k + |\Delta_{\bm{k}}|^2}$. Restricting to the spin singlet channel, we obtain 

\begin{align}
    \Delta_{\bm{k}, \uparrow\downarrow,O(s)} = -\sum_{\bm{k}'}V_{\bm{k}\bm{k}',E(\bm{k})}\Delta_{\bm{k}', \uparrow\downarrow,O(s)}\chi_{\bm{k}'}.
\end{align}

with 

\begin{align}
    V_{\bm{k}\bm{k}',E(\bm{k})} = -V^2\Bigg[ \frac{\omega_{\bm{k} + \bm{k}'}}{(\epsilon_{\bm{k}'} - \epsilon_{\bm{k}} )^2 - \omega^2_{\bm{k} + \bm{k}'}}A(\bm{k} + \bm{k}', \Omega) + \frac{\omega_{\bm{k} - \bm{k}'}}{(\epsilon_{\bm{k}'} - \epsilon_{\bm{k}} )^2 - \omega^2_{\bm{k} - \bm{k}'}}A(\bm{k} - \bm{k}', \Omega)\Bigg],
\end{align}

where $O(s)$ indicates that the gap function is odd in spin, and therefore even in momentum. Comparing with standard phonon-mediated s-wave BCS pairing, this potential has an additional minus sign in front. We therefore instead consider the spin triplet channel, where we find

\begin{align}
    \Delta_{\bm{k}, \uparrow\downarrow,E(s)} = -\sum_{\bm{k}'}V_{\bm{k}\bm{k}',O(\bm{k})}\Delta_{\bm{k}', \uparrow\downarrow,E(s)}\chi_{\bm{k}'},
    \label{GAP_eq}
\end{align}

with

\begin{align}
    V_{\bm{k}\bm{k}',O(\bm{k})} = -V^2\Bigg[ \frac{\omega_{\bm{k} + \bm{k}'}}{(\epsilon_{\bm{k}'} - \epsilon_{\bm{k}} )^2 - \omega^2_{\bm{k} + \bm{k}'}}A(\bm{k} + \bm{k}', \Omega) - \frac{\omega_{\bm{k} - \bm{k}'}}{(\epsilon_{\bm{k}'} - \epsilon_{\bm{k}} )^2 - \omega^2_{\bm{k} - \bm{k}'}}A(\bm{k} - \bm{k}', \Omega)\Bigg].
\end{align}

Considering scattering exactly at the Fermi surface ($\epsilon_{\bm{k}'} = \epsilon_{\bm{k}} = \epsilon_{F}$), the potential simplifies to

\begin{align}
    V_{\bm{k}\bm{k}',O(\bm{k})} = V^2\Bigg[ \frac{1}{\omega_{\bm{k} + \bm{k}'}}A(\bm{k} + \bm{k}', \Omega) - \frac{1}{\omega_{\bm{k} - \bm{k}'}}A(\bm{k} - \bm{k}', \Omega)\Bigg].
\end{align}

When $\bm{k} - \bm{k}'$ is small, i.e.\! when $\bm{k}$ and $\bm{k}'$ are almost parallel and $\Delta_{\bm{k}', \uparrow\downarrow,E(s)}$ has the same sign as $\Delta_{\bm{k}, \uparrow\downarrow,E(s)}$, the second term in the potential dominates and the potential is attractive. When $\bm{k} + \bm{k}'$ is small, i.e.\! when $\bm{k}$ and $\bm{k}'$ are almost anti-parallel and $\Delta_{\bm{k}', \uparrow\downarrow,E(s)}$ has the opposite sign as $\Delta_{\bm{k}, \uparrow\downarrow,E(s)}$, the first term in the potential dominates and the potential is repulsive. In both cases the signs in Eq.\! \eqref{GAP_eq} work out in order to provide a non-trivial solution of the gap equation. The $A$-factor clearly strengthens the interaction, which increases the critical temperature of the superconducting instability.\\

In order to determine the critical temperature, we linearize the gap equation 

\begin{align}
    d_{\bm{k}} = -\sum_{\bm{k}'}V_{\bm{k}\bm{k}',O(\bm{k})}\,d_{\bm{k}'}\,\frac{1}{2|\epsilon_{\bm{k}'}|}\tanh(\frac{|\epsilon_{\bm{k}'}|}{2k_B T_c}),
\end{align}

where we have defined $d_{\bm{k}} = \Delta_{\bm{k}, \uparrow\downarrow,E(s)} = \Delta_{\bm{k}, \downarrow\uparrow,E(s)}$. Following Ref.\! \cite{Sigrist1991}, we write

\begin{align}
    d_{\bm{k}} = -D_0\langle V_{\bm{k}\bm{k}',O(\bm{k})}\,d_{\bm{k}'}\rangle_{\bm{k}', \rm{FS}}\,\int^{\omega_c}_{-\omega_c}\rm{d}\epsilon\, \frac{1}{2|\epsilon|}\tanh(\frac{|\epsilon|}{2k_B T_c}),
\end{align}

where $D_{0}$ is the single-particle density of states at the Fermi level, $\omega_c$ is a cutoff energy for the boson spectrum and $\langle\,\,\,\,\,\,\rangle_{\bm{k}', \rm{FS}}$ is an angular average over the Fermi surface. Assuming $\omega_c/(k_B T_c) >> 1$, we can take

\begin{align}
    \frac{1}{\lambda} = \int^{\omega_c}_{-\omega_c}\rm{d}\epsilon\, \frac{1}{2|\epsilon|}\tanh(\frac{|\epsilon|}{2k_B T_c}) \approx \ln(\frac{1.14\,\omega_c}{k_B T_c}),
\end{align}

implying

\begin{align}
    k_B T_c = 1.14\,\omega_c\,e^{-1/\lambda},
\end{align}

where the dimensionless coupling constant $\lambda$ is the largest eigenvalue of the eigenvalue equation

\begin{align}
    \lambda \, d_{\bm{k}} = -D_0\langle V_{\bm{k}\bm{k}',O(\bm{k})}\,d_{\bm{k}'}\rangle_{\bm{k}', \rm{FS}}.
\end{align}

By picking discrete points on the Fermi surface for $\bm{k}$ and $\bm{k}'$, this equation can be expressed as a matrix eigenvalue problem

\begin{align}
    \lambda \,\bm{d} = M \bm{d},
\end{align}

which can be solved numerically for a given set of model parameters in order to determine $\lambda$ as well as the corresponding eigenvector, which contains information about the structure of the gap function.\\

\section{Material parameters}

In the long-wavelength limit the density of states $D(\epsilon) = \sum_{\bm{k}}\delta(\epsilon - \epsilon_{\bm{k}})$ of the tight binding model is 

\begin{align}
    D(\epsilon) = \frac{N}{4\pi t}.
\end{align}

Taking $t = 0.8\,\rm{eV}$, produces 

\begin{align}
    \frac{D_0}{N} = 1.4 \,\frac{1}{\rm{atom\,Ry}},
\end{align}

which is a typical magnitude for the long-wavelength density of states of metals such as Cu, Al and Au \cite{Burdick1963, Lin2008, Ramchandani1970}. For the Fermi momentum, we take a small value of $k_F\,a = 0.07\,\pi$, which provides us with an approximately circular Fermi surface and makes the results less dependent on the lattice geometry. For the AFMI, we take $J=5 \,\rm{meV}$, $J/K = 2000$ and $s = 1$ \cite{Samuelsen1969, Kobler2006}. The cutoff $\omega_c$ is set to the value at the Brillouine zone boundary $\omega_c = 2szJ$. Finally, the strength of the interfacial coupling is typically reported to be on the order of magnitude of $10\,\rm{meV}$ \cite{Kajiwara2010, Fjaerbu2019}.


\hypersetup{
    colorlinks=true,
    linkcolor=black,    
    citecolor=black,    
    urlcolor=white,      
}
\color{white}

\end{document}